\documentclass[useAMS,usenatbib]{mn2e}

\usepackage{graphicx}

\def\x2{\mbox{$R_{rms}$}}

\title[Magellanic Bridge star clusters and associations]{Bridge over troubled gas: clusters and 
associations under the SMC and LMC tidal stresses\thanks{Based on observations obtained at the 
Southern Astrophysical Research (SOAR) telescope, which is a joint project of the Minist\'erio da 
Ci\^encia, Tecnologia, e Inova\c c\~ao (MCTI) da Rep\'ublica Federativa do Brasil, the US
National Optical Astronomy Observatory (NOAO), the University of North
Carolina at Chapel Hill (UNC) and Michigan State University (MSU).}}

\author[E. Bica et al.]{E. Bica$^1$, B. Santiago$^1$, C. Bonatto$^1$, R. Garcia-Dias$^1$,
L. Kerber$^2$, B. Dias$^{3,4}$, \and B. Barbuy$^3$, E. Balbinot$^5$\\
$^1$ Departamento de Astronomia, Universidade Federal do Rio Grande do Sul, Av. Bento
Gon\c{c}alves 9500\\
Porto Alegre 91501-970, RS, Brazil\\
$^2$ LATO-DCET-UESC, Rodovia Ilh\'eus-Itabuna km 16, Ilh\'eus, Bahia, 45662-000, Brazil\\
$^3$ Universidade de S\~ao Paulo, IAG, Rua do Mat\~ao 1226, S\~ao Paulo 05508-900, Brazil\\
$^4$ Department of Physics, Durham University, South Road, Durham DH1 3LE, UK\\
$^5$ Department of Physics, University of Surrey, Guildford GU2 7XH, UK}

\begin{document}

\pagerange{\pageref{firstpage}--\pageref{lastpage}}

\maketitle

\label{firstpage}

\begin{abstract}
We obtained SOAR telescope B and V photometry of 14 star clusters and 2 associations in
the Bridge tidal structure connecting the LMC and SMC. These objects are used to study
the formation and evolution of star clusters and associations under tidal stresses from
the Clouds. Typical star clusters in the Bridge are not richly populated and have in general
relatively large diameters ($\approx30-35$pc), being larger than Galactic counterparts of
similar age. Ages and other fundamental parameters are determined with field-star decontaminated 
photometry. A self-consistent approach is used to derive parameters for the most-populated sample
cluster NGC\,796 and two young CMD templates built with the remaining Bridge clusters. We find 
that the clusters are not coeval in the Bridge. They range from approximately a few Myr (still 
related to optical HII regions and WISE and Spitzer dust emission measurements) to about 100-200 
Myr. The 
derived distance moduli for the Bridge objects ($1^h56^m<\alpha<2^h28^m$) suggests that the 
Bridge is a structure connecting the LMC far-side in the East to the foreground of the SMC to 
the West. Most of the present clusters are part of the tidal
dwarf candidate D\,1, which is associated with an H\,I overdensity. We find further evidence 
that the studied part of the Bridge is evolving into a tidal dwarf galaxy, decoupling from the 
Bridge.
\end{abstract}


\section{Introduction}
\label{intro}

\citet{Westerlund71} referred to their blue and violet plate survey of stars and 17 star clusters
as an SMC Wing structure. However, they were probing a significant extension of the currently known
Inter-Cloud or Magellanic Bridge of stars and gas connecting the Clouds (\citealt{Westerlund90};
\citealt{BS95}, and references therein). The dynamical evolution of the Clouds has been addressed
by N-body simulations (\citealt{BC05}, \citealt{B+12}, \citealt{Kalil13}) showing that a flat, old 
LMC disk would have evolved to a bar, thick disk and halo, through interactions with the SMC. 

The Magellanic System is extremely rich in extended objects, i.e. star clusters, associations and
emission nebulae. \citet{BS95} and \citet{BSDO99} provided an overall census 
with 1074 extended objects in the SMC, 6659 in the LMC, along with 114 in the Bridge. At that time, a total
of 7847 extended objects were catalogued. Concerning the Bridge census, a borderline at
$\alpha=2^h$ was adopted between the SMC Wing and the Bridge (\citealt{BS95}).
Based on the more recent 
literature and their discoveries, \citet{BBDS08} compiled 9305 extended objects in the Magellanic
System, containing 144 in the Bridge.

With newly discovered clusters and already catalogued ones at the time, \citet{Hodge86} provided
601 star clusters in an SMC census; later, \citet{Hodge88} compiled 2053 clusters in the LMC. The
SMC-LMC Bridge is known since long (\citealt{MM66}), but its stellar content and emission nebulae
have been little by little disclosed (\citealt{Westerlund71}; \citealt{Meaburn86}; \citealt{BS95};
\citealt{BSDO99}). Stellar associations across the entire SMC-LMC connection were discovered by
\citet{IDK90} and \citet{BP92}. These objects are typically extended (with diameters $\ga5\arcmin$
or about 85 pc). Using
CCD photometry, \citet{Grondin90} estimated an age of 100 Myr for the association IDK\,3, projected
at $7\degr$ from the SMC and $14\degr$ from the LMC.

\citet{Demers91} analysed the association IDK\,6 (= ICA\,76), which is the closest one projected
towards the LMC. They obtained an age of 100 Myr and concluded that it should be
spatially located at the LMC far side. \citet{DemBat98} estimated ages of 10-25 Myr for 6 Bridge
associations (10-63 Myr for ICA\,16, closer to the SMC), with distances appearing to vary within
14 kpc. \citet{P+07} derived ages for two relatively rich star clusters at the SMC
Wing/Bridge borderline, obtaining 110 Myr for NGC\,796 (= L\,115 or WG\,9) and 140 Myr for L\,114.
 Very recently, \cite{P+15} carried out a study of 51 clusters in the
outermost eastern region of the SMC. The clusters were observed in the infrared colours
YJK$_s$ within the VISTA Magellanic Clouds (VMC) survey \citep{Cioni11}.
Seven objects are in common between the present work and \cite{P+15}.

The Inter-Cloud region can provide information on how the stellar population of a giant tidal
structure breaks into a stellar density spectrum, from associations down to star clusters. In
turn, this well resolved nearby laboratory can lead us to model and better understand tidal
counterparts in more distant interacting galaxies. The present study has several perspectives:
{\em (i)} using colour-magnitude diagrams (CMDs) to study the radial density profile
(RDP) of clusters and relatively small associations in the Magellanic Bridge; {\em (ii)} provide
the age and density distributions of the different stellar aggregates and infer a scenario for the
cluster and association formation and evolution in the Bridge, under the enduring action of tidal
forces from both the SMC and LMC. In particular, we intend to set constraints to available age and
distance ranges (e.g. \citealt{DemBat98}) using field-star decontamination tools (e.g.
\citealt{BoBi07}; \citealt{BoBi10}), and self-consistent parameter analyses recently applied to
SMC star clusters (\citealt{Dias14}). We recall that our methods are described in previous  work 
dealing with analysis of
young clusters, in particular embedded ones, in the Galaxy (e.g. \citealt{BoBi09}; \citealt{Saurin10}; 
\citealt{Camargo13}). We intend to apply the same methods to clusters in the Clouds.

In Sect.~\ref{sample} we describe the observations and reductions and
collect parameters in the literature that can be useful for comparison and
used as input values
for our analysis. In Sect.~\ref{sCMDs} we present general properties of the sample objects. 
In Sect.~\ref{RDPs}  we discuss cluster structures by means of RDPs 
and present the CMDs.
In Sect.~\ref{groups} we analyse the CMDs of NGC\,796 and two young templates.
In Sect.~\ref{method} we describe the method of analysis and in Sect.~\ref{results} we present the results, 
as well as all clusters individually.   
In Sect.~\ref{discuss} a discussion is given. 
Finally, in Sect.~\ref{conclu} conclusions are drawn.

\section{Observations and reductions}
\label{sample}

We obtained images of 15 fields (together with offset fields) towards the SMC-LMC Bridge, containing 14 
star clusters and 2 associations. Images were taken in B, V and R using the 
Southern Astrophysical Research (SOAR) telescope, equipped with the SOAR Optical Imager
(SOI) detector. The R band 
images were used exclusively for building composed colour images. SOI consists of two $2048 \times 4096$ 
CCDs, each one with two separate controllers. The total FOV is 5.5 arcmin on a side. We adopted a $2\times 2$ 
pixel binning, which yields a $0.154\arcsec\,\rm pixel^{-1}$ scale. Fields offset by 5 arcmin North of each 
cluster position were also imaged to help model the background stellar field. We call these off-cluster fields 
to distinguish them from the on-cluster fields. For each on-cluster and off-cluster fields, we took long and 
short exposures. Two short exposure sets of 20s, 15s, and 15s were taken in B, V and R, respectively. These 
were used for bright stars. As for the long exposures, we took $3 \times 600$s in B, we had $3 \times 200s$ 
in V and $2 \times 200$s in the R filters. 

\begin{figure}
\resizebox{\hsize}{!}{\includegraphics{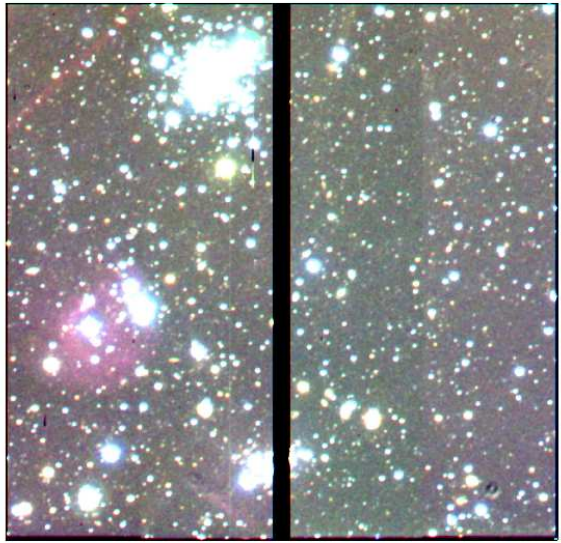}}
\caption{Composite BVR image of BS\,216, centred on the left CCD. NGC\,796 (above)
 and WG\,8 (below) are also
included. North to the top and East to the left.}
\label{fig1}
\end{figure}

Data were taken in service mode throughout several months in 2009 and 2010, as part of SOAR Brazilian time 
share. The observation log is given in Table~\ref{tab1}. We illustrate in Fig.~\ref{fig1} 
the composite BVR field of the BS\,216 pointing, where the two CCDs are seen. This field also sampled the 
clusters NGC\,796 and WG\,8. A gap of $5\arcsec$ occurs between the CCDs. We always have a cluster centred 
on the left CCD. 

Photometric standards (\citealt{Sharpee02}) were observed each night at different airmasses. In the nights 
of January 16th and 17th, 2010, however, the latter fields were too high. We then used standards in the SN 
1987A neighbourhood (\citealt{WS90}). Exposure times for the standards varied in order to allow high S/N 
measurements over the full range of standard mags. From 30 to 50 bias frames were taken by the SOAR resident 
astronomers prior to each observing run. In addition, 20 dome flat images were taken with the 3 filters, 
except for the last two nights, when 15 dome flats were taken. Typical counts per flat exposure were 16-18K 
for B, 17-22K for V, and 18-20K for R. The data reductions were carried out within IRAF, using the {\em mscred} 
and {\em soar} packages. For each night, a single coadded image was created from each set of the run bias 
and flat exposures. Extreme points in the distribution of count values were eliminated and the median count 
value from the remaining frames was stored in the coadded calibration images. These were then visually 
inspected and had their noise statistics cross examined to those of individual exposures. Science images 
were then processed. All exposures, including the bias and flat coadds, were first trimmed according to 
the TRIMSEC parameter in the header. The overscan region of each CCD controller (BIASEC parameter) was 
fitted with a cubic spline and the resulting count level at each column was subtracted from the exposed 
areas. This was done for the science exposures and calibration images. The residual bias image was then 
subtracted from the science exposures and flats. Finally, the exposures were divided by the resulting 
flat images in each filter, in order to account for non-uniform pixel sensitivity. The exposures were 
then separated according to on-cluster or off-cluster, filter and exposure time and then coadded. A median 
filtering was used. The offsets among the exposures to be combined were measured in an automated way, 
by finding the offset value in CCD positions that maximized the number of matched bright stars in each 
exposure relative to the first one in the list.

\subsection{Calibration}
\label{calib} 

The photometric standards were identified in each exposure, also using an automatic finder. It runs a 
basic astrometry using the pixel scale and header parameters and then tries to maximize the number of 
stars from the list of standards, given a positional tolerance. Measurements on these standards were 
also made automatically using a 5 pixel aperture. The IRAF digiphot task PHOT was used. A PSF model 
was built on each of the standard exposures. The PSF candidates were pre-selected in each standard 
field (2 from \citealt{Sharpee02} and 1 from SN1987A) and were found using the same approach as in 
the case of the photometric standards. A circularly symmetric gaussian PSF was fitted to the stars 
in each exposure. The IRAF digiphot PSF task was used for this purpose. A calibration equation of 
the form 
\begin{equation}
\label{eq1}
m=\mu + A\times colour + B\times X + C
\end{equation}
was fitted in separate for 
each night using the standards, where $m$ is the intrinsic mag in the Johnson-Cousins system, $\mu$ 
is the instrumental mag, colour is the standard colour in B-V, $X$ is the airmass, and $A, B$, and 
$C$ are fitting coefficients. In order to prevent low S/N measurements and saturated stars from
contaminating the calibration, a $2\sigma$ clipping criterion (Table~\ref{tab1}) to the list of 
calibrating stars was applied iteratively until convergence. 

\begin{table}
\caption[]{Log of observations}
\label{tab1}
\renewcommand{\tabcolsep}{2.3mm}
\begin{tabular}{ccccl}
\hline\hline
    Date    &  Band    &   N    &$\sigma$ & Target\\
\hline
2009-12-17  &    B     &    29  &  0.068  & BS\,216, NGC\,796, WG\,8 \\
2009-12-17  &    V     &    29  &  0.060  & BS\,216, NGC\,796, WG\,8 \\
2009-12-18  &    B     &    31  &  0.062  & WG\,11  \\
2009-12-18  &    V     &    33  &  0.138  & WG\,11  \\
2009-12-26  &    B     &    31  &  0.080  & WG\,13  \\
2009-12-26  &    V     &    33  &  0.150  & WG\,13  \\
2010-01-16  &    B     &    18  &  0.028  & BS\,245  \\
2010-01-16  &    V     &    19  &  0.024  & BS\,245  \\
2010-01-17  &    B     &    22  &  0.064  & BS\,240  \\
2010-01-17  &    V     &    24  &  0.120  & BS\,240  \\
2010-08-30  &    B     &    21  &  0.056  & BS\,223  \\
2010-08-30  &    V     &    18  &  0.023  & BS\,223  \\
2010-09-10  &    B     &    23  &  0.090  & BS\,226, BS\,225, WG\,15  \\
2010-09-10  &    V     &    22  &  0.035  & BS\,226, BS\,225, WG\,15  \\
2010-09-30  &    B     &    15  &  0.023  & BS\,232  \\
2010-09-30  &    V     &    21  &  0.050  & BS\,232  \\
2010-10-01  &    B     &    27  &  0.047  & BS\,235, BS\,233  \\
2010-10-01  &    V     &    29  &  0.103  & BS\,235, BS\,233 \\
2010-10-17  &    B     &    18  &  0.045  & BS\,239   \\
2010-10-17  &    V     &    20  &  0.085  & BS\,239  \\
2010-11-13  &    B     &    29  &  0.184  & BBDS\,4  \\
2010-11-13  &    V     &    32  &  0.131  & BBDS\,4   \\
\hline
\end{tabular}
\begin{list}{Table Notes.}
\item N is the number of standard star measurements; $\sigma$ for clipping criterion
\end{list}
\end{table}

\begin{figure}
\resizebox{\hsize}{!}{\includegraphics{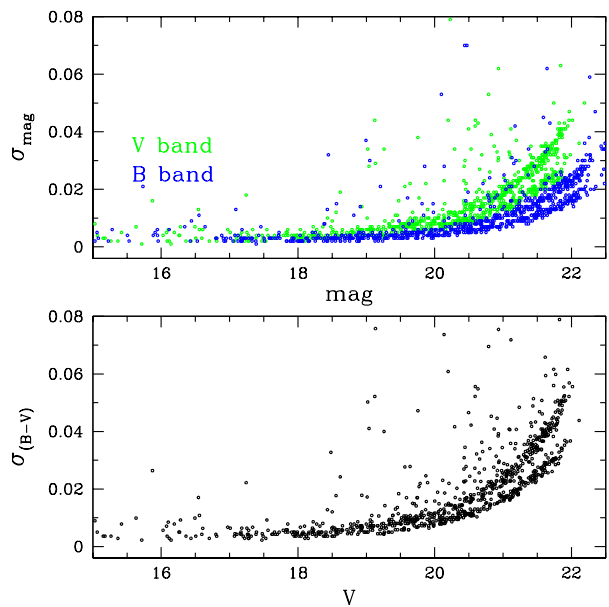}}
\caption{Photometric uncertainties in V and B magnitudes (upper panel) 
and in (B-V) colour (lower panel} for NGC\,796.
\label{fig2}
\end{figure}

Similarly, the IRAF {\em digiphot} package was used to source detection and photometry in 
the on-cluster and off-cluster coadds. A threshold of 5 times the sky fluctuation was used 
for source detection on the V band coadds. This source table was used to perform aperture 
photometry in the B and V filters. The mean FWHM in each coadd was used for aperture 
photometry. A circularly symmetric Gaussian PSF model was also built on each on-cluster and 
off-cluster coadds. In each image the PSFs of the 40 brightest unsaturated stars were visually 
inspected in order to select the best ones to build the PSF models. The ALLSTAR task was used 
to compute x and y centres, sky values, and mags. performing fits of the PSF models to groups 
of stars in each coadd. Parameters as the fitting and PSF radii were taken based on the mean 
FWHM for each image, as described in the DAOPHOT manual. Each star center was recomputed in 
every fit iteration, and so was the sky value, considered as a single value for each group 
of stars in the fitting. The calibration equation for each night was then applied to the science 
fields. The calibrated magnitude and intrinsic colours were computed simultaneously 
for each star. The instrumental colour was used as an initial guess in the calibration equation, 
yielding an initial value of the calibrated magnitude. A new colour was then computed and the process 
reiterated until all colours converged to 0.01 mag. Uncertainties in the calibrated magnitudes were 
derived by propagating the uncertainties in the instrumental magnitudes, and calibration parameters.
In order to illustrate these values, we present the final uncertainties in magnitudes and colour for NGC\,796 in Fig. \ref{fig2}.

\begin{figure}
\resizebox{\hsize}{!}{\includegraphics{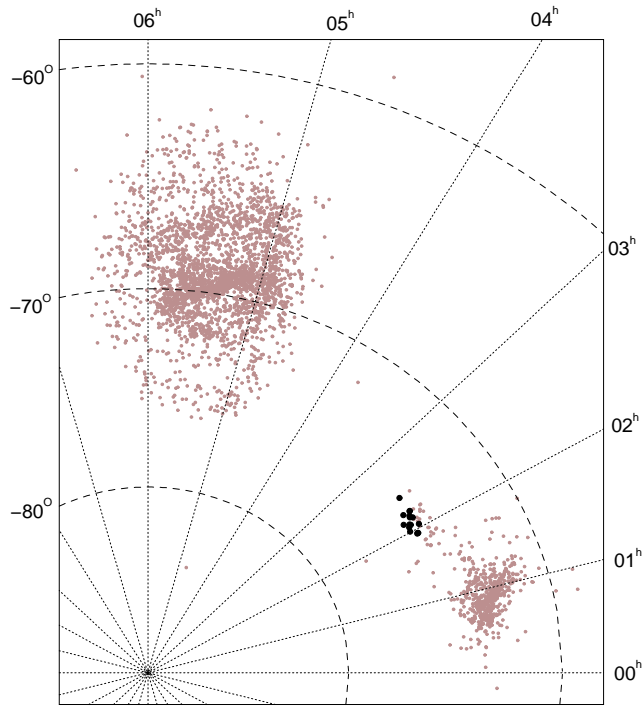}}
\caption{Location of the sample clusters and associations (black points) with respect to 
the LMC and SMC cluster positions in the catalogue by \citep{BBDS08}).}
\label{fig3}
\end{figure}

\begin{figure*}
\resizebox{\hsize}{!}{\includegraphics{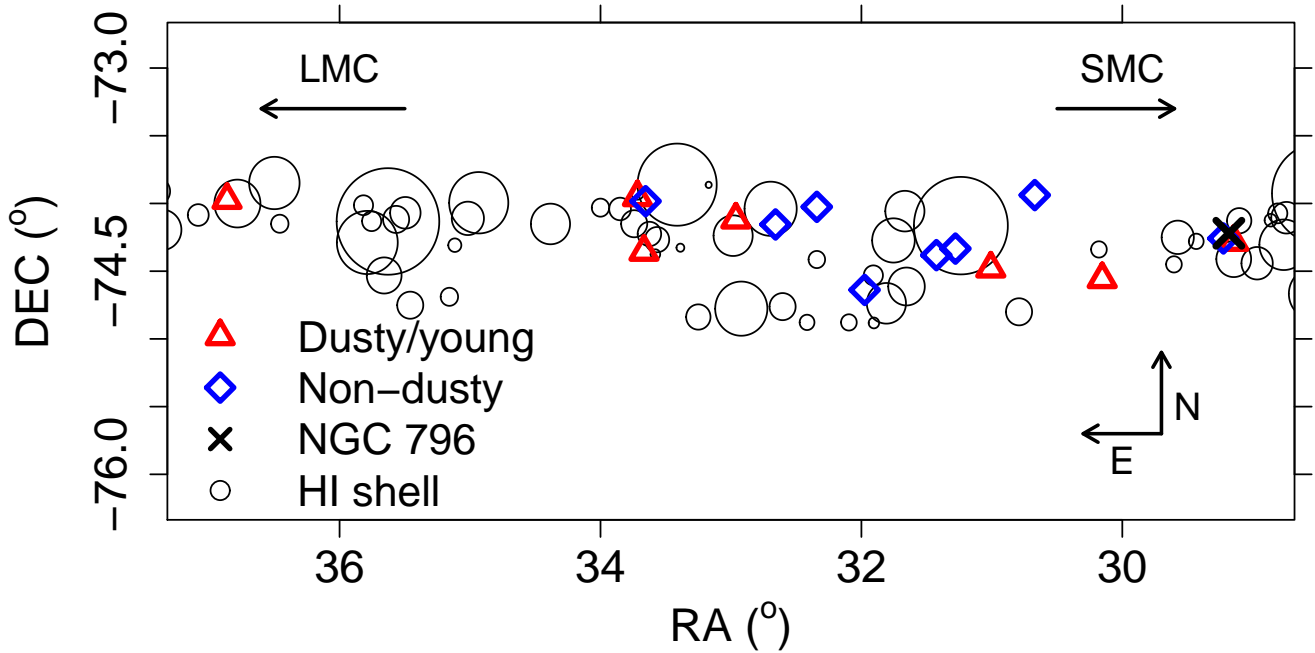}}
\caption{Location of the sample clusters and H\,I shells with respect to the LMC and SMC,
where the data points are from \citet{BBDS08}. North to the top and East to the left.}
\label{fig4}
\end{figure*}

\section{General properties of the sample objects}
\label{sCMDs}

Fourteen  star clusters and two associations could be extracted 
from the SOAR images, as shown in Table~\ref{tab3}, 
for which the density classes (C, A, AC, CA) are from  \citet{BS95}. 
The acronyms in Table~\ref{tab3} are  L58 (\citealt{Lindsay58}), 
L61 (\citealt{Lindsay61}), WG(\citealt{Westerlund71}), 
ESO (\citealt{Lauberts82}), ICA (\citealt{BP92}), BS (\citealt{BS95}),
BBDS (\citealt{BBDS08}).

We show in Fig. \ref{fig3} the positions of the present sample clusters with respect
to the overall clusters in the \cite{BBDS08}, which traces the present 
structure with respect to the Bridge, LMC and SMC.

\begin{figure*}
\resizebox{\hsize}{!}{\includegraphics{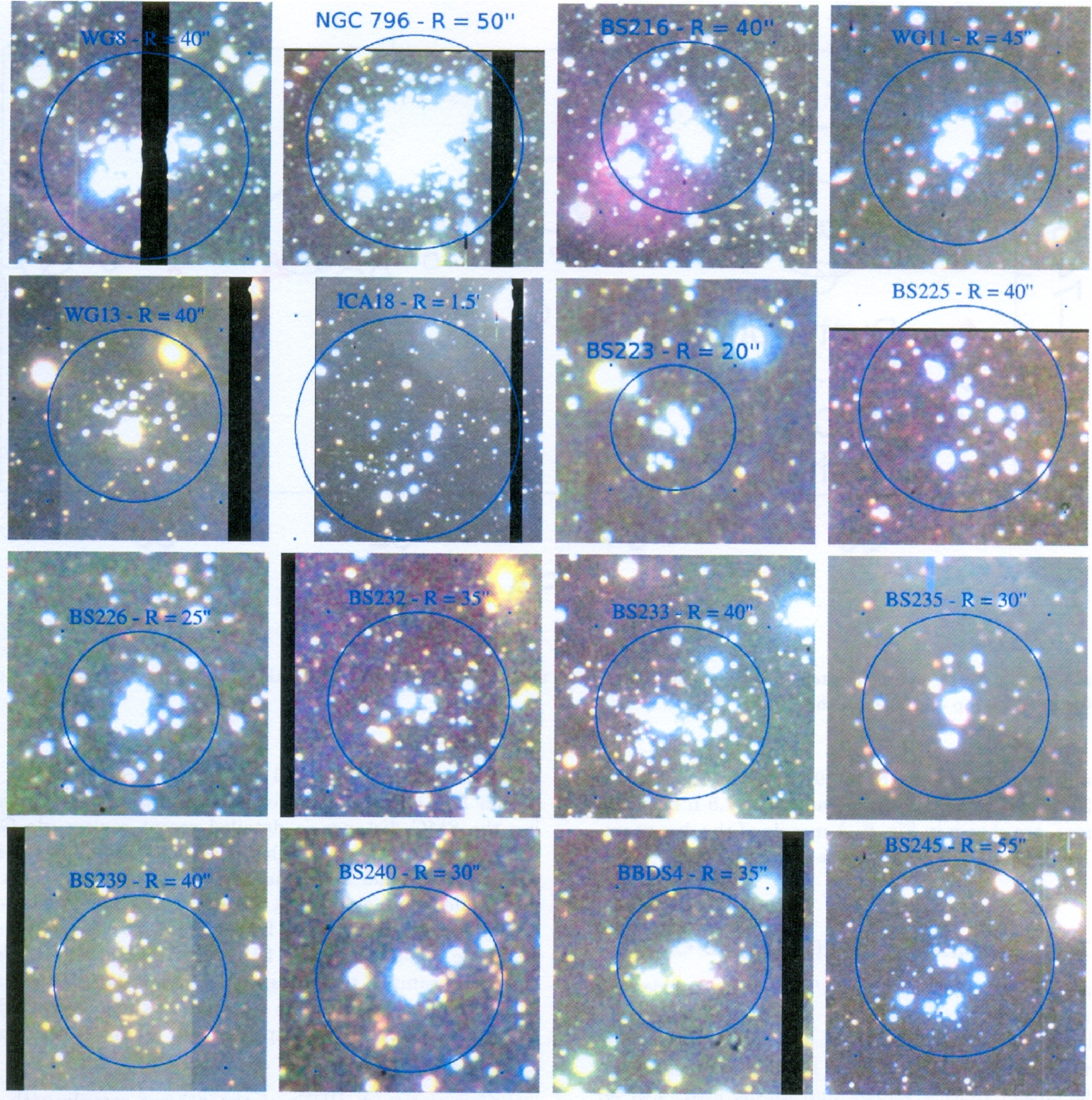}}
\caption{Composite BVR images of the sample targets centred on the coordinates in Table~\ref{tab3}.
Each panel is $1\times1\arcmin^2$. Circles show the bulk of the cluster stars. North to the top and 
East to the left.}
\label{fig5}
\end{figure*}

In Fig. \ref{fig4} the present clusters are shown in the corresponding
Magellanic Bridge sector, together with the H\,I shells in the area. The directions to
both Clouds are shown, as well as our classification of dusty and non-dusty
clusters (see Table~\ref{tab3} and Sect.~\ref{groups}). Coordinates and sizes of
clusters and H\,I shells are from \cite{BBDS08} and references therein. Dusty clusters basically correlate angularly 
with H\,I shells.  Fig.~\ref{fig5} shows a mosaic of BVR images of the objects, built with the present
observations. The mosaic shows a range of stellar densities, from extremely compact to loose
ones. The clusters BBDS\,4 and BS\,216 appear to be associated to  gas emission in optical 
images.

The compact cluster WG\,15 is projected close to the edge of the association ICA\,18. Too few
stars could be retrieved by the extraction routine in WG\,15, so we analyse the ensemble of 
ICA\,18, including WG\,15. 

We indicate in Table~\ref{tab2} the foreground reddening values predicted by \citet{Schle98} 
and \citet{Schlafly11} for the present sample cluster directions, taking three of them for their 
strategic positions along the Bridge. We indicate the cluster values closer to the SMC Wing, farthest 
away from the SMC in the Bridge, and one third of the way between both.

\begin{table}
\caption[]{Reddening for objects across the Bridge from HI columns}
\label{tab2}
\renewcommand{\tabcolsep}{1.1mm}
\begin{tabular}{lccccc}
\hline\hline
Cluster & $\alpha$ & $\delta$ & $A_V^{(1)}$ & $A_V^{(2)}$ & Note\\
        & (h:m:s)  & (\degr:\arcmin:\arcsec) & (mag) & (mag) \\
(1) & (2) & (3) & (4) & (5) & (6)\\
\hline
BS\,216 & 1:56:54 & -74:15:28 & 0.158 & 0.131 & close to SMC Wing \\
BS\,245 & 2:27:27 & -73:58:23 & 0.166 & 0.138 & far in the Bridge \\
BS\,233 & 2:10:38 & -74:09:20 & 0.181 & 0.222 & one third away \\
\hline
\end{tabular}
\begin{list}{Table Notes.}
\item (1): \citet{Schle98}; (2): \citet{Schlafly11}.
\end{list}
\end{table}

Some sample objects seem to be asymmetric in the stellar distribution, such as BS\,216 and 
especially WG\,13, in the sense that there occurs a deficiency of stars in a hemisphere, with 
respect to the bright central star (Fig.~\ref{fig5}). Asymmetry effects do not appear to be 
an exclusive trait of young Magellanic clusters (see e.g. the Galactic proto-clusters in 
\citealt{CBB11} and \citealt{Camargo15}), but they are
 rare in large collections of CCD cluster
images of the SMC and LMC (\citealt{Pietrzynski98}; \citealt{Pietrzynski00}). The effect 
is probably related to star formation and evolution under two sources of tidal stresses. 
 Asymmetrical tidal effects might e.g. preferentially erode the molecular 
clouds exposing molecular cores, or affect the early dynamical evolution by exposing stellar 
cores as a consequence of  evaporation.

\begin{table*}
\caption[]{The Bridge sample}
\label{tab3}
\renewcommand{\tabcolsep}{4.5mm}
\begin{tabular}{lcccccll}
\hline\hline
Designation & $\alpha$ & $\delta$ & class & D & d  & Part of & Dust\\
            & (h:m:s)  & (\degr:\arcmin:\arcsec) & & (\arcmin) & (\arcmin) \\
(1) & (2) & (3) & (4) & (5) & (6) & (7) & (8)\\
\hline
 WG\,8                            & 1:56:35 & -74:17:00 & AC & 1.00 & 0.50 &  in BS\,215 & Y \\
 NGC\,796,L\,115,WG\,9,ESO\,30SC6 & 1:56:44 & -74:13:10 &  C & 1.20 & 1.10 &  in BS\,215 & N \\
 BS\,216$^\ddagger$               & 1:56:54 & -74:15:28 &  C & 0.40 & 0.40 &  in BS\,217 & Y \\
 WG\,11$^\ddagger$                & 2:00:37 & -74:33:29 &  C & 0.70 & 0.55 & & Y \\
 WG\,13$^\ddagger$                & 2:02:41 & -73:56:21 &  C & 1.10 & 0.85 & & N \\
 ICA\,18$^{\dagger\ddagger}$       & 2:07:54 & -74:38:17 &  A & 3.20 & 2.30 & & N \\
 BS\,223$^\ddagger$               & 2:04:02 & -74:28:48 &  C & 0.35 & 0.35 &  in BS\,224 & Y \\
 BS\,225                          & 2:05:07 & -74:20:02 &  A & 1.10 & 0.75 & & N \\
 BS\,226$^\ddagger$               & 2:05:42 & -74:22:53 &  C & 0.40 & 0.35 &  in ICA9 & N \\
 BS\,232$^\ddagger$               & 2:09:22 & -74:01:31 & CA & 0.80 & 0.80 &  in BS\,231 & N \\
 BS\,233$^\ddagger$               & 2:10:38 & -74:09:20 & CA & 1.20 & 1.10 &  in DEMS169 & N \\
 BS\,235$^\ddagger$               & 2:11:51 & -74:07:08 &  C & 0.60 & 0.40 & & Y \\
 BS\,239$^\ddagger$               & 2:14:37 & -73:59:00 & CA & 1.20 & 0.65 &  in ICA\,34 & N \\
 BS\,240$^\ddagger$               & 2:14:52 & -73:57:10 &  C & 0.30 & 0.30 &  in BD34 & Y \\
 BBDS\,4$^\ddagger$               & 2:14:40 & -74:21:24 &  C & 0.30 & 0.20 &  in L61-593 & Y \\
 BS\,245$^\ddagger$               & 2:27:27 & -73:58:23 & CA & 1.10 & 0.90 &  in ICA\,44 & Y \\
\hline
\end{tabular}
\begin{list}{Table Notes.}
\item $\dagger$ The compact cluster WG\,15 (2:07:40, -74:37:45 , C, 0.35\arcmin) is located at the
edge of association ICA\,18. By columns: (1) object designation(s), (2) and (3) J2000.0 equatorial 
coordinates, (4) density class, (5) and (6) major and minor diameters, (7) part of larger object, 
in general an association. (8) evidence of embedded cluster from SPITZER and/or WISE dust emission: 
Y(es) or N(o). Information from \citet{BBDS08} and references therein. $\ddagger$ Cluster centred 
on the CCD.
\end{list}
\end{table*}

\begin{figure}
\resizebox{\hsize}{!}{\includegraphics{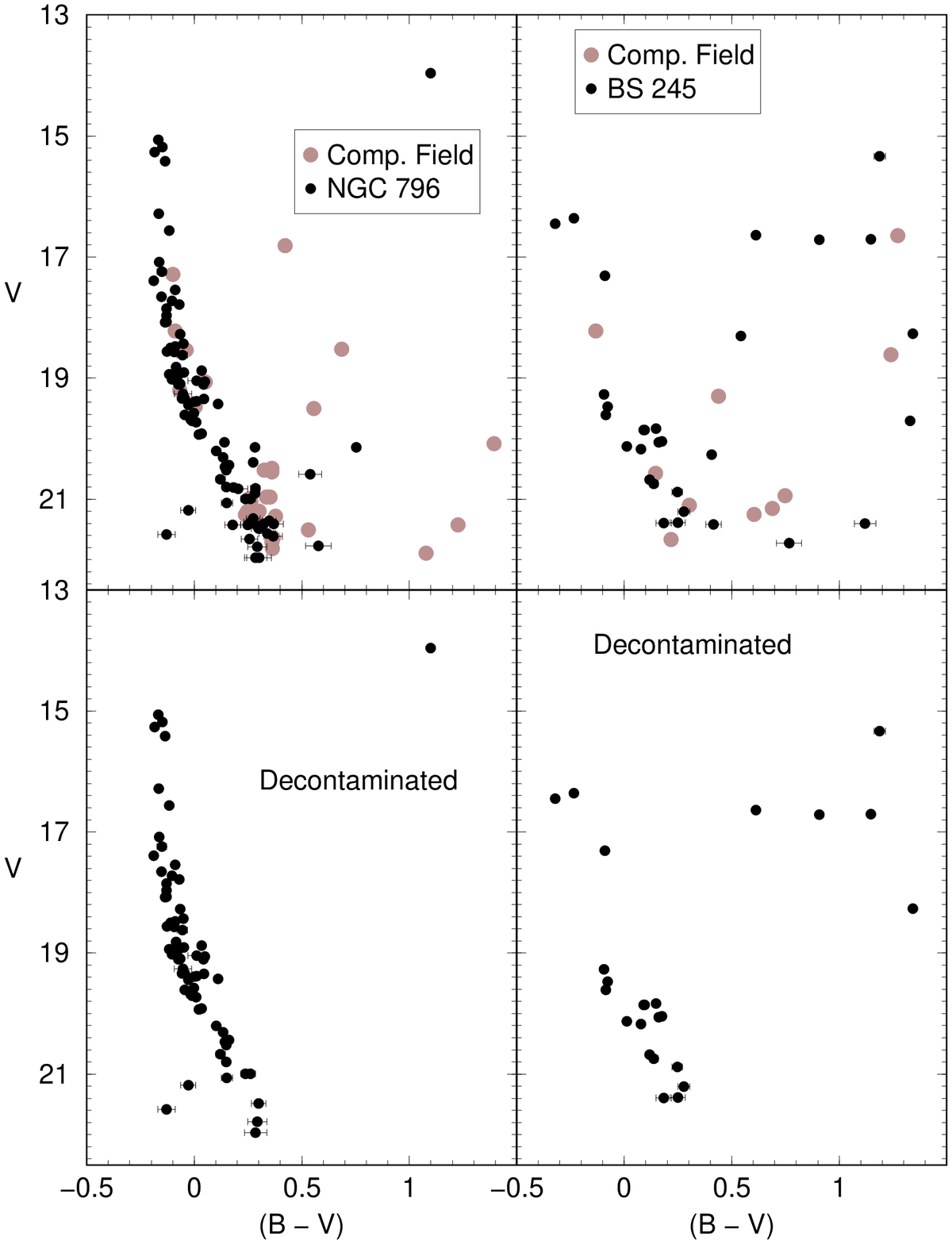}}
\caption{Field-star decontamination examples for NGC\,796 and BS\,245.}
\label{fig6}
\end{figure}

Although field star contamination is not critical in the intercloud region, we carried out a
statistical field subtraction following \citet{BoBi07}. Examples of the decontamination 
procedure for NGC\,796 near the SMC Wing, and BS\,245 quite far in the Bridge at 
$\Delta\alpha$ = 1.5 h from the SMC center, are shown in Fig.~\ref{fig6}. 
A mosaic of the decontaminated CMDs of the whole sample is given in Fig.~\ref{fig8}.
 

\section{Radial density profiles and CMDs}
\label{RDPs}

\begin{figure*}
\resizebox{\hsize}{!}{\includegraphics{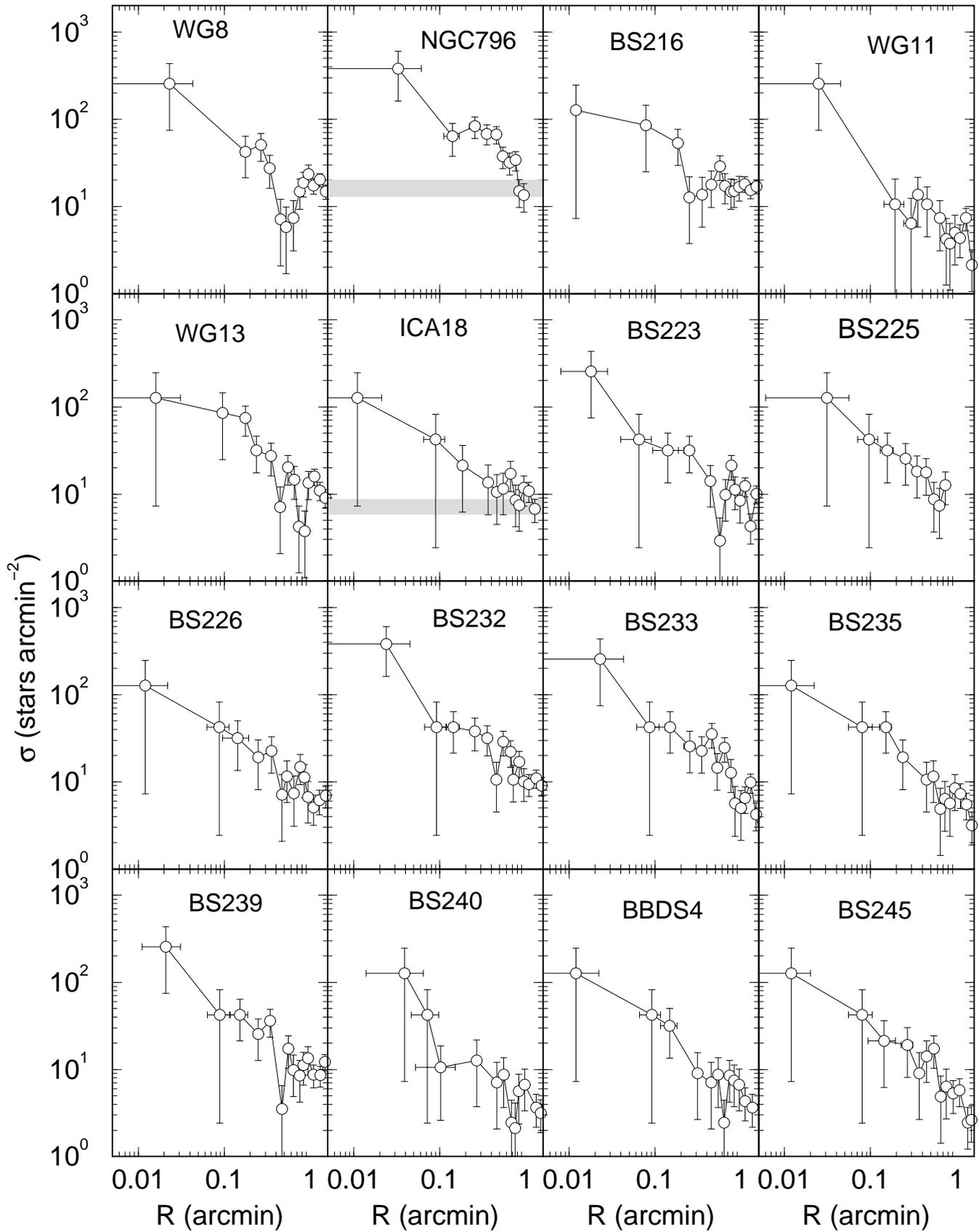}}
\caption{Stellar density distributions of clusters and associations. Grey stripes illustrate
background level and uncertainty for NGC\,796 and ICA\,18.}
\label{fig7}
\end{figure*}

We show in Fig.~\ref{fig7}  the resulting mosaic of Radial density profiles (RDPs). The RDPs 
show in general significant central stellar densities and radial extensions which we measure by means 
of R$_{RDP}$ (e.g. \citealt{CBB11}), where the cluster density equals that of the field. The individual 
shapes of the profiles indicate core/halo structures immersed on a rather shallow background. For all 
objects in Table~\ref{tab3} we searched for automated optimized centres with maximum density, within a 
radius $r=10\arcsec$. The background level and uncertainty are estimated based on the star counts 
and fluctuation measured on the surrounding area beyond the cluster radius (Fig.~\ref{fig7}). 
Usually, we used all the available area outside the cluster to estimate the background level and
perform the field decontamination. Since the projected area of the targets vary and the CCD field 
of view is fixed, the comparison field changes with every case. However, we point out that, in all cases, 
the comparison field area is larger than those of the clusters.

 Examples are
given in Fig.~\ref{fig7}, in which we adopted $1\,\sigma$ values for the background uncertainty.
This allows to estimate the RDP extent, basically defined as the radius where background and cluster
profile meet (e.g. \cite{CBB11}). We illustrate in Fig.~\ref{fig6} the background determination 
for NGC\,796 and BS\,245. For NGC\,796, we obtain $R_{RDP}\approx0.8\arcmin$, and for BS\,245, 
$R_{RDP}\approx1\arcmin$. Regarding the SMC distance, a detailed recent review (\citealt{DGB15})
came up with the distance modulus value of $(m-M)_O=18.96\pm0.02$ mag (62\,kpc). Note, however, that
depth effects are significant (\cite{Crowl01}). For simplicity, we assume 62\,kpc. In this case,
the absolute radii would be $\sim14.4$\,pc and $\sim18$\,pc. Considering the RDPs of the whole 
sample, the Bridge clusters are larger than most of the young clusters in the Galactic disk (e.g. 
\citealt{Saurin12}). So, they appear to be inflated with respect to Galactic counterparts. 

Comparing all RDPs of the present sample with the respective absolute
radii of the Galactic star cluster catalogue of \citet{Kh13} in Fig.~\ref{fig9},
it is clear that the Bridge objects occupy the large-radii wing of the distribution. The sample
of \citet{Kh13} contains primarily clusters old enough to be trimmed by the Milky Way tidal
field. One possible conclusion is that Bridge clusters are formed as relatively large structures.
Note that the present section assumes a SMC distance of 62\,kpc. Given that the Bridge connects
the SMC with the LMC and that implies a gradient in distance corresponding to a distance modulus
variation of at leas 0.5\,mag, we also consider the effects on cluster size for a shorter distance 
of $\sim$ 40\,kpc for the Bridge clusters (Sect. 8). They still occupy the large-size wing in 
Fig.~\ref{fig9}.  

We propose that the Bridge clusters are inflated  as a consequence of tidal fields in their 
formation and early evolution  under  the SMC and LMC tidal stresses. Interestingly, clusters
with structural asymmetries also present a break towards the core in the RDPs 
(Fig.~\ref{fig7}).

\section{NGC\,796 and Young CMD templates}
\label{groups}

As observed in a 4m class telescope, most sample clusters seem to 
be relatively poorly-populated, but we recall that we are dealing with young Magellanic clusters, 
so they may be intrinsically more massive. Thus, to mitigate the rather-low statistics, we decided 
to create templates based on their CMDs. The dusty clusters (Table~\ref{tab3}) form a young group 
with 5 entries: BS\,223, BS\,235, BS240, BS\,245, BBDS\,4. Although dusty, BS\,240 has not been included 
in the template because its CMD resulted very poor.

The remaining clusters and associations, ICA\,18, BS\,239, BS\,233, BS\,232, WG\,13, BS\,216, BS\,225 
and BS\,226, are not dusty (Table~\ref{tab3}). They all have CMDs very similar to that of NGC\,796, 
which might suggest a coeval star-formation event. All but one of the dusty clusters share their 
same spatial location on the Bridge with non-dusty clusters (see Fig. \ref{fig4}). The groups are built 
by applying, when necessary, small shifts in magnitude ($\sim0.03$) and, to a lesser extent, colour 
($\sim0.01$). The reason for this is to obtain statistically significant CMDs with which to derive parameters such
as age, reddening and distance. Then, the individual CMDs can be matched to the respective template ones
to derive the individual parameters. At this point, we have 3 populous CMDs suitable for a more
detailed analysis (Fig.~\ref{fig10}):  (i) NGC\,796 itself, (ii) a dusty template, and (iii) a dust-free 
one, very similar to NGC\,796.

\begin{figure*}
\resizebox{\hsize}{!}{\includegraphics{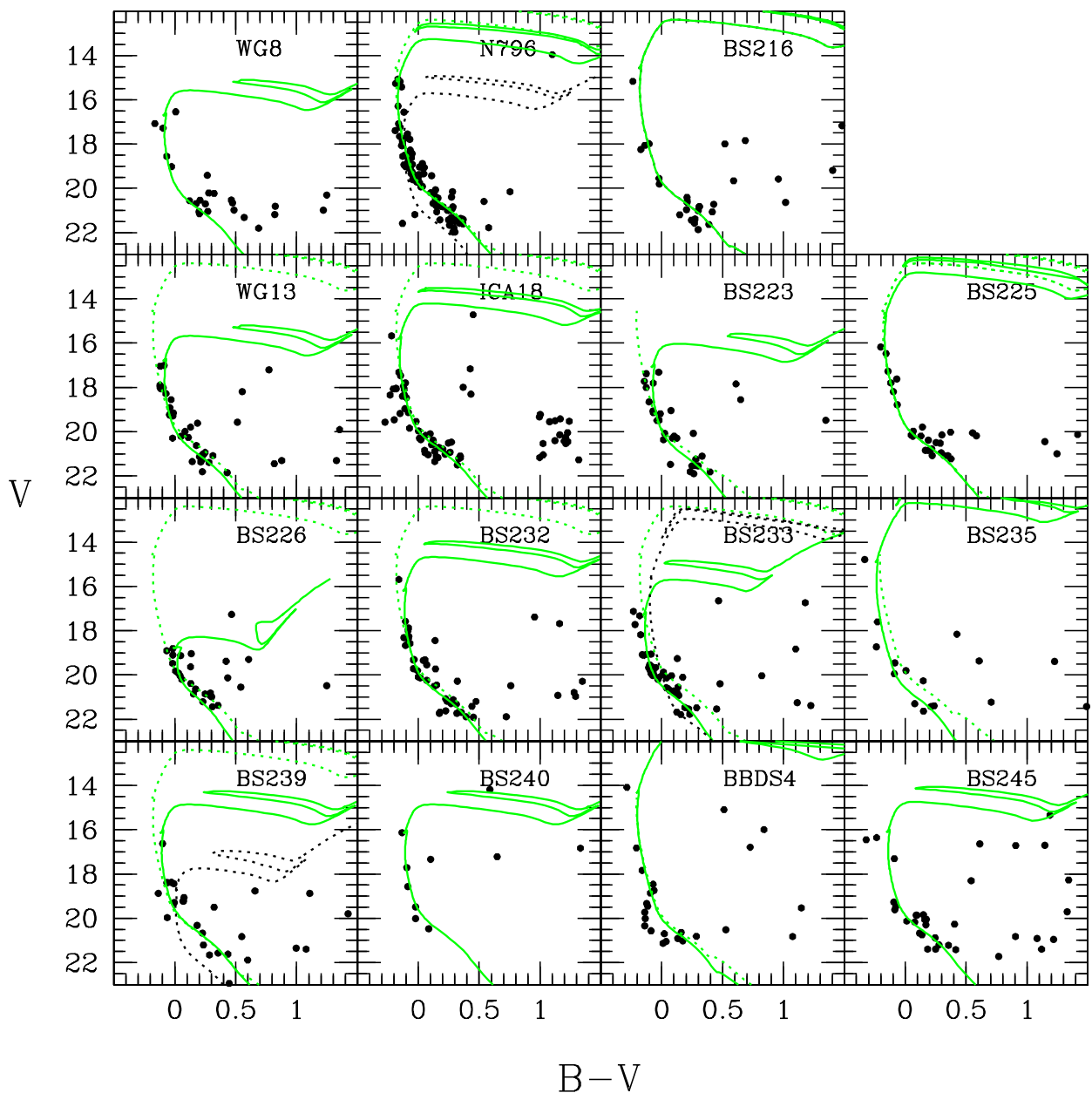}}
\caption{Visual isochrone fittings to the decontaminated CMDs all clusters and associations 
(green solid lines), except for NGC\,796, where a numerical-statistical method was used to 
find the best solution (see Sect. \ref{method}). The physical parameters used for these fits 
are presented in Table~\ref{tab5}. For comparison purposes, the template isochrone fittings 
(green dotted lines) and the physical solutions proposed by \citet{P+07} (NGC\,796) and 
\citet{P+15} (BS\,233, BS\,239 and BBDS\,4) are also presented (black dotted lines).
The size of photometric uncertainties vs. magnitude and colour are illustrated in Fig.\ref{fig2}.}
\label{fig8}
\end{figure*}

\begin{figure}
\resizebox{\hsize}{!}{\includegraphics{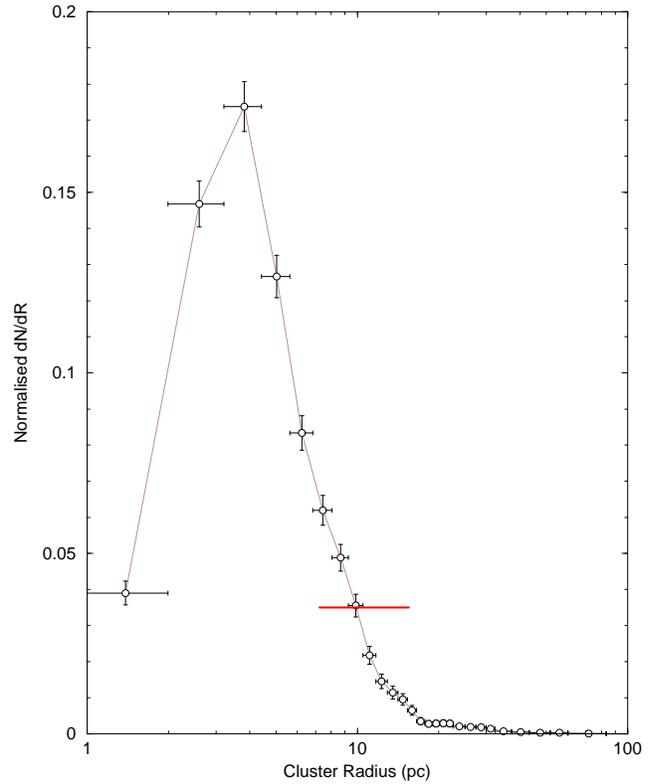}}
\caption{Comparison of the radius of the present sample objects (red) with Galactic star 
clusters (\citealt{Kh13}), assuming a SMC distance modulus of 62\,kpc (\citealt{DGB15}).}
\label{fig9}
\end{figure}

\begin{figure}
\resizebox{\hsize}{!}{\includegraphics{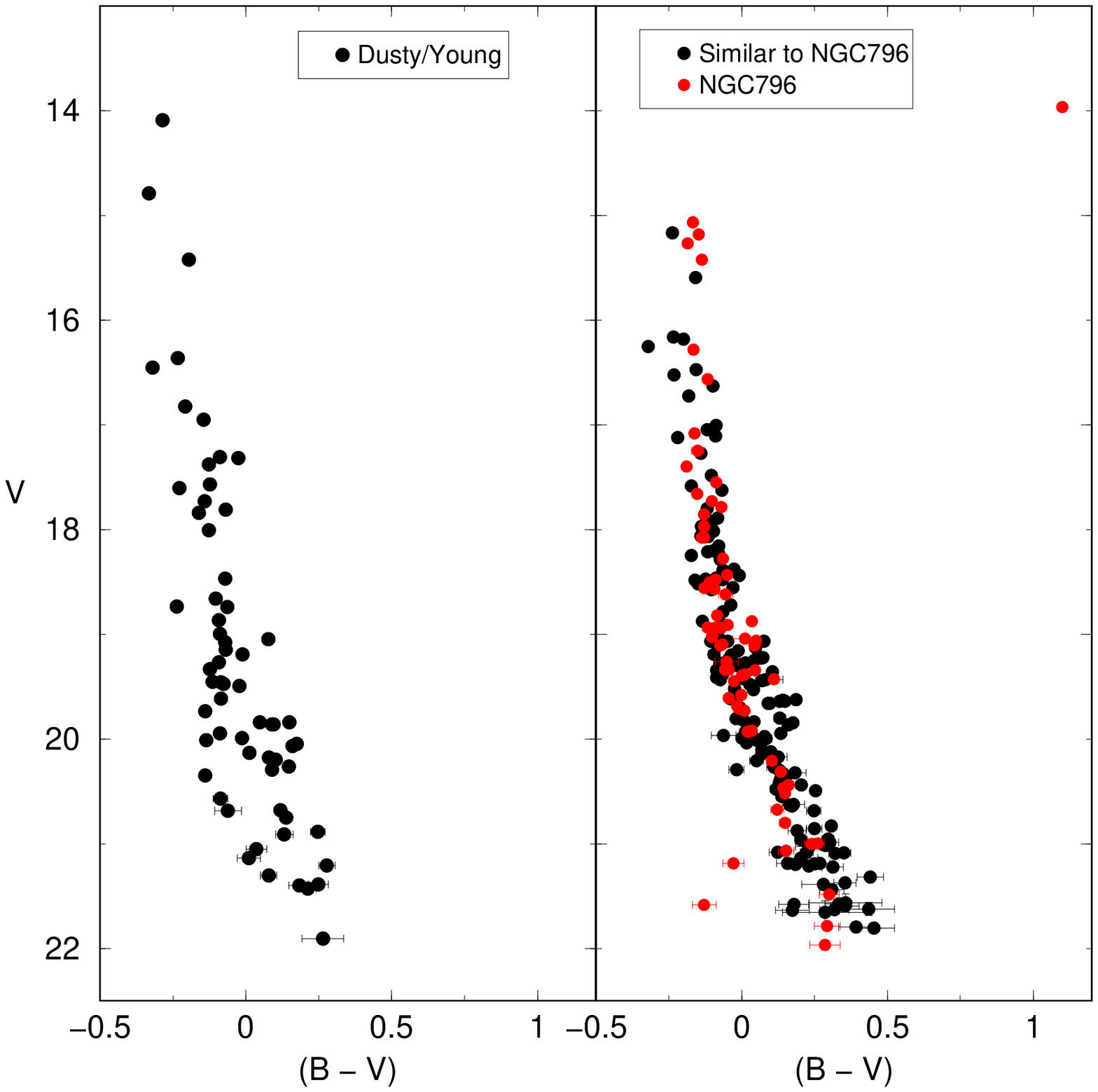}}
\caption{Photometric templates. Left: dusty CMDs. Right: dust-free CMDs compared to the
populous NGC\,796.}
\label{fig10}
\end{figure}


\section{Method of analysis}
\label{method}

\begin{figure}
\resizebox{\hsize}{!}{\includegraphics{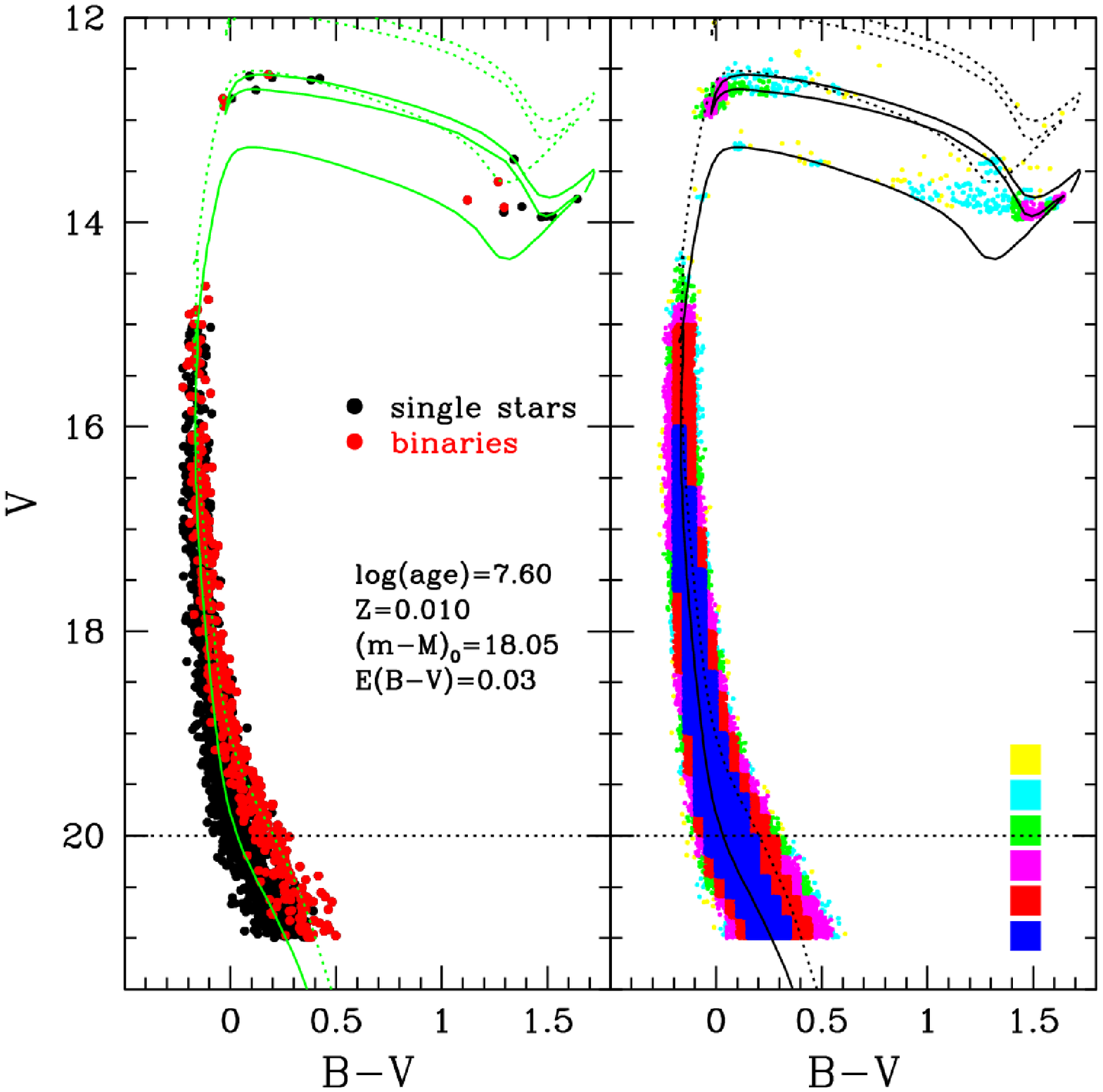}}
\caption{Generation of synthetic CMDs. Left: an example of synthetic CMD generated with 2000 stars. The adopted 
isochrone (green solid line) and its physical parameters are presented in the panel. Right: the same as in the 
left panel, but for 10000 stars. The colours in this case are following the density of points, which means the 
probability to generate stars in that CMD position. The isochrone for equal mass binaries is also shown in both 
panels (dotted lines). We adopted a power-law mass function following the Salpeter slope and a binary fraction of 30\%.}
\label{fig11}
\end{figure}

In order to recover the age, metallicity, distance modulus and reddening for NGC\,796 and for the two 
templates we applied a numerical-statistical isochrone fitting. This method was initially developed to 
derive physical parameters for LMC clusters observed with HST \citep{K02, KS05, KSB07}, but more recently 
it was used to analyse open clusters in the near-infrared \citep{A12} and SMC clusters observed with SOAR 
under very similar conditions with respect to the data presented here \citep{Dias14}.
 In short, it is based on objective comparisons between observed CMDs and synthetic ones using a likelihood statistics.  
Those models that maximize this statistics are identified as the best ones, 
and then they are used to constrain the stellar cluster physical parameters.
Fig. \ref{fig11} illustrates the generation of a synthetic CMD that was compared with the observed CMDs in this work. 
Essentially the probability of a star to belong to a specific model is measured by the density of points in its CMD. 
So the likelihood of this model will be proportional to the product of these probabilities for the observed stars.
       
To avoid local or biased solutions we have explored a wide range of values 
and a regular grid of models in the space parameter, 
virtually rising all possible solutions that can be considered among the best ones.
Typically we use $\sim 10^{4}$ combinations in log($\tau$/yr), $Z$, 
$(m-M)_{0}$ and $E(B-V)$ centred in an initial guess obtained by a visual isochrone fit.
In terms of distance modulus, we investigated solutions from 17.70 to 19.30, by far covering 
the most acceptable results for the SMC distances and its complex geometry \citep{DGB15}, 
including the large line-of-sight depth for the SMC eastern side ($\sim 23$ kpc) recently 
found by \cite{Nidever+13}. 
Concerning the age, metallicity and reddening, we searched for solutions in the following ranges: 
$6.60 <$ log($\tau$/yr)$ < 10.00$ (steps of 0.05); $0.0001 <$ Z $< 0.0020$ (steps of $\sim$ 0.002); 
$0.00 <$ $E(B-V)$ $< 0.30$ (steps of 0.01).
Once again, these ranges are sufficiently wide to cover any result concerning the SMC or the Bridge, 
like the ones present in the SMC age-metallicity determinations for stellar clusters 
\citep{Piatti11, Parisi+14} and field stellar populations \citep{Carrera+08, Rubele+15}, 
in the analysis of Bridge field stars \citep{Nidever+13,Skowron+14}, and reddening 
maps \citep{Burstein82,Schle98}.

Since the uncertainties in the physical parameters are computed using synthetic CMDs that are reproducing the observed ones, the spread of stars and the stochastic effect due to the photometric uncertainties and the limit number of observed stars are naturally incorporated in our results.
For further details we recommend the reader to consult the 
aforementioned works.

\begin{figure}
\resizebox{\hsize}{!}{\includegraphics{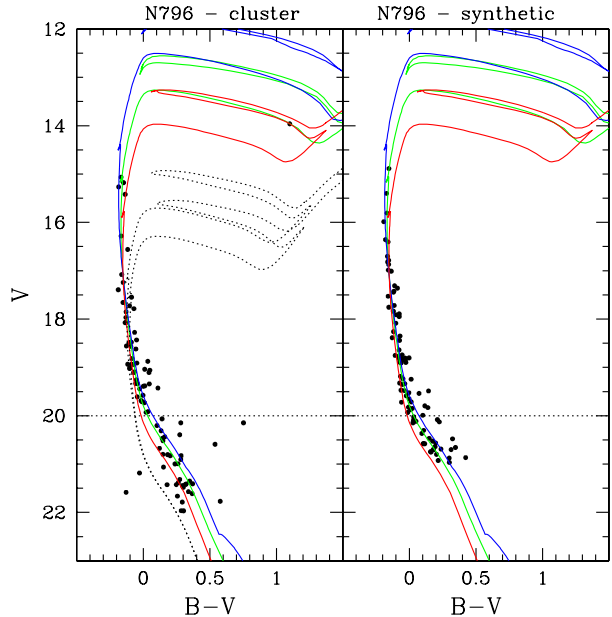}}
\caption{Best isochrone fittings for NGC\,796.
Left panel: observed stars and three isochrones with the parameters found in Table~\ref{tab4}. The 
different solid lines correspond to the central solution (green), the young/metal-rich solution (blue) and the 
old/metal-poor solution (red). The physical solution proposed by \citet{P+07} is also presented (dotted line). 
Right panel: synthetic CMD generated with the parameters found for the best solution.}
\label{fig12}
\end{figure}

\begin{figure}
\resizebox{\hsize}{!}{\includegraphics{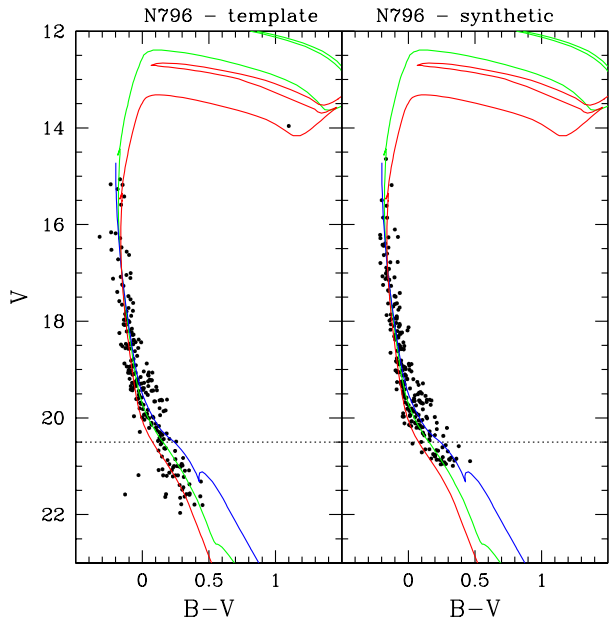}}
\caption{The same as in Fig.~\ref{fig12} but for the NGC\,796 template.}
\label{fig13}
\end{figure}

\begin{figure}
\resizebox{\hsize}{!}{\includegraphics{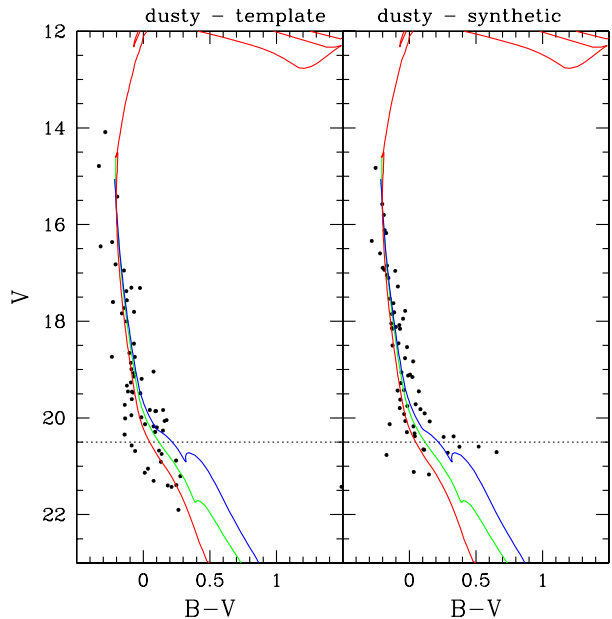}}
\caption{The same as in Fig.~\ref{fig12} but for the dusty/young template.}
\label{fig14}
\end{figure}

\begin{table*}
\caption{Self-consistent physical parameters derived by us using our numerical-statistical isochrone fitting.}
\label{tab4}
\centering
\begin{tabular}{lccccccc}
\hline 
Name & log($\tau/yr$) & Age(Myr) & Z & [Fe/H] & (m-M)$_{0}$ & d(kpc) &
E(B-V)  \\
\hline
NGC\,796 & $7.63\pm0.19$ & $42 ^{+24}_{-15}$ & $0.010\pm0.005$ & $-0.3^{+0.2}_{-0.3}$ & $18.04\pm0.28$ & $40.6\pm1.1$& $0.03\pm0.02$\\
NGC\,796 - template & $7.41\pm0.24$ & $25 ^{+18}_{-11}$ & $0.010\pm0.005$ & $-0.3^{+0.2}_{-0.3}$ & $18.01\pm0.26$ & $40.1\pm1.1$& $0.04\pm0.02$\\
Dusty - template & $7.22\pm0.14$ & $16 ^{+6}_{-5}$ & $0.006\pm0.004$ & $-0.5^{+0.2}_{-0.5}$ & $17.96\pm0.26$ & $39.2\pm1.1$& $0.04\pm0.03$\\
\hline
\end{tabular}
\end{table*}

\section{Results}
\label{results}

Based on the results of the previous sections, 
we use the CMD templates to estimate
parameters for individual clusters by computing magnitude 
and colour differences with respect to the templates 
(including NGC\,796).

Table \ref{tab5} reports physical parameters derived by a
visual isochrone fitting method, as shown in Fig. \ref{fig8}.
The uncertainties presented in this table correspond to the range of acceptable solutions 
based on a visual inspection. Certainly these values are conservative, and they can be used as 
an upper limit.
On the other hand, a lower limit for the uncertainties in these determinations can 
be found in Table~\ref{tab4}, where the random 
uncertainties for the numerical-statistical method are presented.

Almost all clusters have ages from log(age)=7.30 (BBDS\,4) 
to log(age)=8.30 (BS\,233). The only exception is BS\,226, the 
oldest cluster in our sample (log(age)=8.95).
Another interesting result is related to the the distance determinations:
all clusters are located at distance modulus between 17.95 (BBDS\,4) and 18.40 
(WG\,13 and BS\,233), therefore even closer than the canonical LMC value 
of $(m-M)_0=18.50$ mag \citep{dGWB14}.    

We have NGC\,796 in common with \cite{P+07} (Washington photometry 
with ESO 1.54 Danish telescope) and
BS\,226, BS\,232, BS\,233, BS\,235, BS\,239, BS\,240 and BBDS\,4 in common
with \cite{P+15} (Near-infrared photometry with VISTA). 
A comparison with their results leads to the following conclusions. 

\cite{P+15} imposes a metallicity of Z=0.003, and a distance
modulus of $(m-M)_0$ = 18.90 mag, and adopted the E(B-V) values from several sources.
The age is then derived from visual isochrones fittings to the CMDs. We instead
fit all parameters, such that differences are pointed out below.

{\bf NGC\,796}

A comparison of the present fit for NGC\,796 (L115) and that by \cite{P+07}
 is shown in Figs. \ref{fig12} and \ref{fig13}.
They derived visually an age of log(age)=8.05-8.20
for NGC 796, assuming that it is at a distance modulus
of  $(m-M)_0$ = 18.77, a metallicity of Z=0.003 
and a reddening E(B-V)=0.03, from maps by \cite{Burstein82}
and \cite{Schle98}.
It is clear that the solution by \cite{P+07} does not fit the
brightest stars of the cluster (their fit corresponds to the
dotted lines). An inspection of data used by  \cite{P+07}
indicates that the images used were saturated for the bright stars,
and no short exposures are mentioned.
Therefore we are confident that NGC 796 is younger than reported by
 \cite{P+07}, with an age of log(age)$\sim 7.6$,
and that it is located at a short distance 
($(m-M)_0 \sim 18.04$, $d\sim 41$ kpc). 

{\bf BS\,226}, {\bf BS\,232}, {\bf BS\,235} and {\bf  BS\,240}
\cite{P+15} classify these cluster as ``possible non-cluster'' or ``not recognized''.
However, our deeper photometry indicates them as clusters, especially for the main
sequence distributions shown in Fig.~\ref{fig8} for BS\,226 and BS\,232. The remaining
two have less-populated main sequences.

{\bf BS\,233}
For BS\,233 \cite{P+15} report an age
of log(age)=7.3 yr, having assumed a reddening of E(B-V)=0.15,
and a distance modulus of $(m-M)_0 = 18.90$, and metallicity Z=0.003,
 whereas the present results give E(B-V)=0.0 and
log(age)=8.3 yr, together with a metallicity of Z=0.001. 
Figure~\ref{fig8} shows that the present solution
fits the data more satisfactorily.

{\bf BS\,239}
Fig.~\ref{fig8} shows that the present results of
E(B-V)=0.04, Z=0.010 and log(age/yr)=8.0 fit better the data
than the \cite{P+15} solution of 
E(B-V)=0.11, Z=0.003 and log(age/yr)=8.5.

{\bf BBDS\,4}
Fig. \ref{fig8} shows that both solutions from the present work of
E(B-V)=0.01, Z=0.004 and log(age/yr)=7.3 and those from
the \cite{P+15} solution of E(B-V)=0.08, Z=0.003 and 
log(age/yr)=9.1 fit the data reasonably well. This cluster is clearly
embedded in an H\,II region.

{\bf  BS\,245}
Fig.~\ref{fig8} shows some spread, but a main sequence seems to be
confirmed.

\begin{table*}
\caption{Physical parameters (and uncertainties) derived by us using visual 
isochrone fits.}
\label{tab5}
\renewcommand{\tabcolsep}{3.5mm}
\begin{tabular}{lcccc}
\hline 
Name & log($\tau/yr$) & Z & (m-M)$_{0}$ & E(B-V)  \\
\hline
WG\,8 & $8.10\pm0.25$ & $0.010\pm0.006$ & $18.30\pm0.30$ & $0.05\pm0.04$ \\
BS\,216 & $7.60\pm0.30$  & $0.010\pm0.006$ & $18.00\pm0.30$ & $0.04\pm0.04$ \\
WG\,13 & $8.10\pm0.25$ & $0.010\pm0.006$ & $18.40\pm0.30$ & $0.05\pm0.04$ \\
ICA\,18 & $7.80\pm0.30$ & $0.010\pm0.006$ & $18.20\pm0.30$ &$0.02\pm0.02$ \\
BS\,223 & $8.20\pm0.25$ & $0.008\pm0.004$ & $18.40\pm0.25$ &$0.05\pm0.04$ \\
BS\,225 & $7.50\pm0.30$ & $0.010\pm0.006$ & $18.00\pm0.30$ &$0.04\pm0.04$ \\
BS\,226 & $8.95\pm0.15$ & $0.002^{+0.002}_{-0.001}$ & $18.00\pm0.20$ & $0.04\pm0.04$ \\
BS\,232 & $7.90\pm0.30$ & $0.008\pm0.006$ & $18.20\pm0.30$ & $0.05\pm0.04$ \\
BS\,233 & $8.30\pm0.20$ & $0.001^{+0.001}_{-0.0006}$ & $18.00\pm0.25$ & $0.00^{+0.02}_{-0.00}$ \\
BS\,235 & $7.40\pm0.30$ & $0.002^{+0.002}_{-0.001}$ & $18.20\pm0.30$ & $0.00^{+0.02}_{-0.00}$ \\
BS\,239 & $8.00\pm0.25$ & $0.010\pm0.006$ & $18.00\pm0.30$ & $0.04\pm0.04$ \\
BS\,240 & $8.00\pm0.25$ & $0.010\pm0.006$ & $18.00\pm0.30$ & $0.04\pm0.04$ \\
BBDS\,4 & $7.30\pm0.30$ & $0.004^{+0.004}_{-0.003}$ & $17.95\pm0.30$ & $0.01^{+0.02}_{-0.00}$ \\
BS\,245 & $8.00\pm0.25$ & $0.008\pm0.006$ & $18.00\pm0.30$ & $0.01^{+0.02}_{-0.00}$ \\
\hline
\end{tabular}
\end{table*}

A comparison of parameters in Table~\ref{tab5} (considering the uncertainties) with
the recent values by \citet{P+15} shows the following: {\em(i)} a non-negligible difference
concerning E(B-V), with ours consistent with the external values quoted in Table~\ref{tab2};
{\em(ii)}
(ii) regarding ages,  \citet{P+15} give a younger value for BS 233, but older values for BS 239 and, more significantly, for BBDS 4. 
Although their solution do not fit well the red main sequence stars in our CMDs, 
the discrepancies in age for these last two clusters can be partially explained by
 the absence of the brightest main sequence stars in their CMD analysis;
 {\em(iii)} the metallicities are comparable.

\section{Discussions}
\label{discuss}

The present work shows evidence that the Bridge clusters have sizes 
probably modulated at birth by tidal forces of the LMC and SMC. 
The present sample is projected in the sky nearer to the less-massive SMC galaxy, 
as indicated by the centroids of the LMC $\alpha=05^h31^m00^s$,
$\delta=-69\degr22\arcmin48\arcsec$ and SMC $\alpha=00^h53^m29^s$,
$\delta=-73\degr07\arcmin48\arcsec$ (\citealt{BBDS08}), and the present
sample coordinates in Table~\ref{tab3}.

Compared to a diagnostic-diagram of dynamical evolution for young Galactic clusters 
(\citealt{BoBi10}; \citealt{Saurin12}), the present Bridge clusters (and the two
associations) have sizes larger than the bulk of embedded clusters in the Galaxy. 
However, they are considerably smaller than Trumpler\,37 and  Bochum\,1, which have 
already evolved into associations.

\begin{figure}
\resizebox{\hsize}{!}{\includegraphics{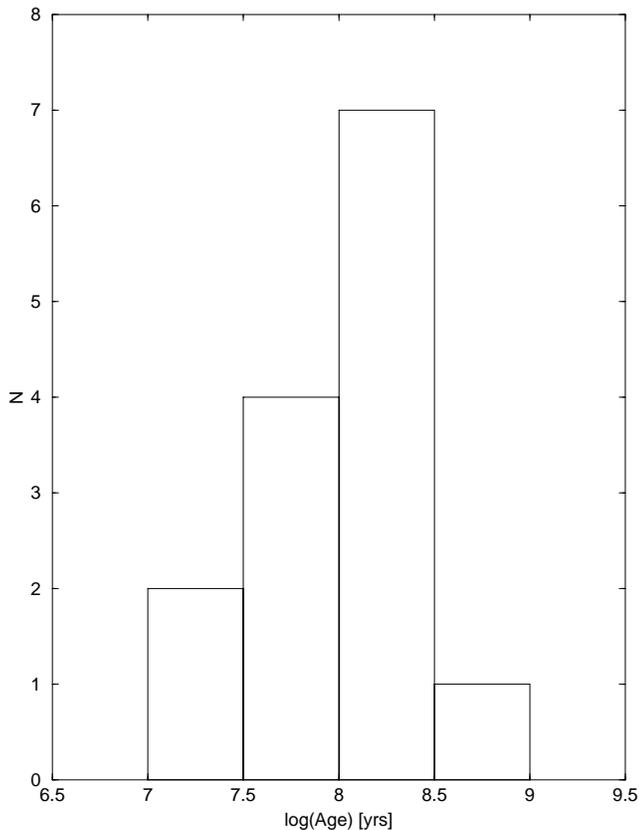}}
\caption{Age histogram for the sample. Note that age uncertainties (Table~\ref{tab5})
are smaller than the bin widths.}
\label{fig15}
\end{figure}

In the last decades much work has been done on depth effects 
of the SMC. 
\citet{Crowl01} used 12 intermediate age and old star clusters to 
find a triaxial distribution of 1:2:4 ($\alpha$:$\delta$:depth), 
with SMC depth effects from 6 to 12\,kpc. 
The 21 cm HI line surveys showed that the SMC, and especially the 
Wing, appear to be split into two velocity components 
(e.g. \citealt{Mat84}; \citealt{MFV86}), with counterparts in Cepheid 
velocities. 
The SMC would be splitted into a Mini-Magellanic Cloud and the 
SMC Remnant, as a consequence of the interactions with the LMC 
(e.g. \citealt{MuFu80}; \citealt{MFV88}; \citealt{GarHaw91}). 
Although these studies propose that the SMC would extend beyond 
its tidal radius, based on Cepheids in the infrared, \citet{Welch87} 
claimed that the SMC should be essentially contained in its 
tidal radius.


We remark that the distance results are a key issue. 
The present clusters appear to lie 
at a distance of $\sim40$\,kpc, thus $\sim20$\,kpc closer 
than the SMC and $\sim10$\,kpc than the LMC. These results 
would not fit the scenario where the Bridge is a simple 
connection between both Clouds, as implied by models like in
 \cite{B+12}. Thus, our work discloses the presence of a 
foreground structure that is projected near the edge of the 
SMC wing. We conclude that the Bridge, having arisen in the far side of the LMC 
(\citealt{Demers91}), may be indeed a tidal arm. 
Recently, \citet{SS15} studied the 
SMC structures reaching the Bridge as well by means of Cepheids. 
Among the results, they find that Cepheids occur in front of the 
SMC tilted plane. \citet{Nidever+13} found evidence
that the stellar component detected at $\sim55$\,kpc is a 
tidally stripped component from 
the SMC 200\,Myr ago following a close encounter with the LMC. 

A deeper insight on the Bridge formation, structure and evolution 
requires a deeper understanding of the LMC and SMC interaction history 
over the last Gyrs, e.g. \citet{Bekki08}, 
\citet{Besla07}, and \cite{Kallivayalil13}.

In addition, one cannot rule out an intermediate-age component in 
the Bridge. For instance, \cite{BCN13} found that ages of 
the RGB and AGB stars in the central Bridge region are likely to range
from $\sim 400$ Myr to 5 Gyr, implying that these stars were drawn into 
the Bridge at the tidal event and did not form in situ. 
The Bridge may not have been formed only by tidal 
stripping, but probably involved also more complex phenomena. Figure~\ref{fig15}
shows the age histogram from Table~\ref{tab5}. Although the small number of objects,
there is a clear dominance towards $\log(Age/years)\sim8.2$. This is a hint to the 
intrinsic star formation history (SFH) in the Bridge. Interestingly, this peak basically
coincides with the last interaction between the LMC and SMC (\citealt{BC05}). We
conclude that the SFH in the Bridge may be mixed with older populations, as described
above.

From the analysis of HI features, \cite{M+04} find complex kinematics inside the Bridge 
and measure $\overline{V_{LSR}} \approx  170 km\,s^{-1}$ in our cluster region. 
\cite{BW04} found $\overline{V_{LSR}} \approx 270 km\,s^{-1}$ for the LMC, 
$\sim160\,km\,s^{-1}$ for the SMC, and $\sim215\,km\,s^{-1}$ for the Bridge. 
Considering the relative velocity between the South-West objects (present sample) in 
the Bridge and that of the LMC since the last interaction between the Clouds \citep{BC05}, we obtain a 
distance difference of $\sim10$kpc, in agreement with the cluster distances in this 
work. Thus, the tidal dwarf candidate D\,1 (\citealt{BS95}) together with its concentration of clusters
and associations are consistent with the observed H\,I kinematics in the area.

\section{Concluding remarks}
\label{conclu}

In the present study we derive reddening, metallicity, age, distance modulus 
and radial density profiles for 14 star clusters and 2 associations of the LMC-SMC Bridge. 
All clusters are found to be quite young, with ages $7.3<\log(\tau/yr)<8.3$ 
(except for BS\,226 with $\log(\tau/yr))=8.95$), 
with reddening in the range $0.01<E(B-V)<0.05$ mag. 
Their metallicities vary in the range $0.001<Z<0.010$,
basically encompassing the values previously found for the young 
populations in the SMC and LMC. 
Regarding reddening, we recall that the total-to-selective absorption
in the SMC indicates R = 2.7, lower than the average Galactic ratio 
(see e.g. \cite{Bouchet85}). 
Although some very young sample clusters have dust emission detected
in Spitzer and/or WISE (Table 3), the detected  reddening values
can be essentially explained from the Galactic foreground 
(Tables 4 and 5). 
Consequently, the dust embedding the clusters does not appear to be 
much effective in absorbing optical light.

\cite{BS95} detected 3 probable dwarf galaxies forming along the LMC-SMC Bridge. 
Differently of the early Universe dwarf galaxies related to stars 
and dark matter, the ones found by \cite{BS95}  in the Bridge appear to be 
tidal products of the latest LMC-SMC interaction. 
Most of the present clusters are part of the tidal dwarf galaxy 
candidate D\,1, which is associated with an HI overdensity, similarly
to D\,2 and D\,3. 
Recently, \cite{B+15} and \cite{K+15} discovered 9 halo dwarf galaxies, 
part of them projected on the surroundings of the Magellanic Clouds 
and the Stream. 
In this context, the young tidal dwarf galaxies populate as 
well the Lynden-Bell plane \citep{LB76}, also called
Vast Polar Structure (VPOS), as recently revised and updated by
\cite{PPK12}. However, the conclusive association to such a gigantic structure
would require kinematical and orbital data for the newly-found dwarf galaxies \citep{K+15}.
Recently, an effort in this direction was the determination of distance and 
heliocentric velocity of the Reticulum\,II ultra-faint dwarf galaxy (\citealt{Simon15}). 
However, the transversal velocity is yet needed for a diagnosis. We suggest that the 
VPOS might as well contain tidal remains of past interactions of the Clouds.

Tidal dwarfs have been detected and described by models (e.g. 
Mirabel et al. 1992; Barnes \& Hernquist 1992), especially for blobs 
in tidal arms as far as 100kpc from the Antennae. 
Such tidal dwarf candidates have already been detected in the Bridge connecting the
LMC and SMC by \cite{BS95} and \cite{BBDS08}. 
Recently, \citet{Ploe15} (and references therein)
presented detailed numerical simulations of the evolution of 
tidal dwarf galaxies, investigating the decoupling of tidal dwarfs 
from the tidal arm.

\section*{Acknowledgements}
 We thank an anonymous referee for important comments and suggestions.
Partial financial support for this research comes from CNPq
 and PRONEX-FAPERGS/CNPq (Brazil). 
BB and BD acknowledge partial financial support from
FAPESP, CNPq, CAPES, and the LACEGAL project.
LK, BB, and BD also thank CAPES/CNPq for their financial support with
the PROCAD project number 552236/2011-0.


\end{document}